\documentclass[iop,numberedappendix,twocolappendix,appendixfloats]{emulateapj}

\usepackage{epsfig}
\usepackage{graphicx}
\usepackage[T1]{fontenc}
\usepackage{natbib}
\usepackage{fmtcount}
\usepackage{rotating}
\usepackage{color}
\usepackage{ulem}
\usepackage{soul,xcolor}

\newcommand{\grs}{GRS 1915+105}
\newcommand{\nustar}{NuSTAR}
\newcommand{\qpo}{QPOs}

\begin{document}

\title{Investigating the Connection Between Quasi Periodic Oscillations 
and Spectral Components  with {\nustar} Data of {\grs}}

\author{Anjali Rao Jassal{\thanks {Email: anjali@prl.res.in}}}
\author{Santosh V. Vadawale}
\author{Mithun N. P. S}
\affil{Physical Research Laboratory, Ahmedabad- 380009, India }

\author{Ranjeev Misra}
\affil{Inter-University Center for Astronomy and Astrophysics, Post Bag 4,
Ganeshkhind, Pune 411007, India}

\begin{abstract}

The low frequency quasi periodic oscillations (QPOs) are commonly observed 
during hard states of black hole binaries. Several studies have established
various observational/empirical correlations between spectral parameters 
and QPO properties, indicating a close link between the two.
However, the exact mechanism of generation of QPO is not yet well understood.
In this paper, we present our attempts to comprehend the connection
between the spectral components and the low frequency QPO observed in 
\grs{} using the data from \nustar{}. Detailed 
spectral modeling as well as the presence of the low frequency QPO and 
its energy dependence during this observation have been reported by
\citet{miller} and \citet{zhang} respectively. We investigate 
the compatibility of the spectral model and energy dependence of the QPO 
by simulating light curves in various energy bands for small variation of the
spectral parameters. 
The basic concept here is to establish connection, if any, between the QPO 
and the variation of either a spectral component or a specific parameter, which 
in turn can shed some light on the origin of the QPO.
We begin with the best fit spectral model of \citet{miller} 
and simulate the light curve by varying the spectral parameter at
frequencies close to the observed QPO
frequency in order to generate the simulated QPO. Further we simulate similar 
light curves in various energy bands in order to reproduce the observed
energy dependence of RMS amplitude of the QPO. We find that the observed
trend of increasing RMS amplitude with energy can be reproduced qualitatively
if the spectral index is assumed to be varying with the phases of the QPO. 
Variation of any other spectral parameter does not reproduce the observed 
energy dependence.


\end{abstract}


\section{Introduction}
Accreting black hole binaries show variability on a wide range of timescales 
from milli-seconds to months and years. While the long term variability 
typically arises due to the transient nature or the state transitions, the
short term (sub-second) variability is generally attributed to the 
processes in the inner regions of accretion disk. 
A perplexing feature of such a
short term variability is the presence of quasi periodic oscillations,
characterized by a narrow peak superimposed over a broad band 
noise in the power density spectrum (PDS).
\qpo{} have been observed in a frequency range of less than Hz to 
hundreds of Hz \citep{rm06} and they are generally classified as low frequency
\qpo{} (LF\qpo{}, 0.1-30 Hz) and high frequency \qpo{} (HF\qpo{}, 40-450 Hz). 
Frequency of HFQPOs does not vary with a sizeable change in
luminosity suggesting that it depends 
on the fundamental system parameters such as black hole mass and spin 
\citep{rm06,rem06b}.

It is evident that the LF\qpo{} are not linked with the orbital motion of 
the material in the disk because the observed frequencies correspond to the far
outer regions of the accretion disk. 
The properties of these QPOs (e.g. low frequency break, centroid frequency, 
rms amplitude) are strongly coupled with the changes in spectrum and 
luminosity \citep[see][]{muno99,sobczak,vig03,li14}, suggesting that these
QPOs originate in the same inner region of the accretion disk.
Despite the knowledge of empirical correlations of QPO 
properties with spectral parameters \citep[e.g.][]{sobczak, stiele}, the 
exact physics is not very well understood. 
There are several models to explain the origin and properties of LF\qpo{}. 
In one of the most promising models, the LFQPO originates as a result of 
relativistic Lense-Thirring (LT) precession of the hot inner flow 
\citep{stella,wagoner,schnit,ingdone12}. The 
model successfully explains the variation of frequencies of QPO centroid 
and the low frequency break in terms of variation in the truncation radius of 
the inner disk. 
Another model based on the accretion ejection instability 
suggests that a QPO is produced by a spiral pattern rotating at a frequency 
of a few tenths of the Keplerian frequency at the inner edge of the disk 
\citep{tagger,varniere,rodriguez02}. However, according to the shock 
oscillation model, wherein a shock is formed by a centrifugal barrier in 
a region of lower viscosity, LFQPOs are originated due to oscillation
of the post-shock region \citep{molteni,charabarti,garain}.

Most of the available QPO models explain the frequencies and some of 
the observed correlations along with other timing and spectral properties.
However, in many cases, the models are silent about the energy dependence 
of the QPO or the exact mechanism of origin of the QPO. 
Observing a QPO as a direct oscillation of one or more 
spectral parameter could provide much deeper insights in the modulation 
mechanism and thus origin of the QPO.  
In this context, here we attempt to investigate the observed energy 
dependence of the LFQPO in \grs{} in terms of modulation of the 
spectral parameters.

\grs{} is one of the most enigmatic black hole binary system famous for its
relativistic jets exhibiting superluminal motion \citep{mirabel94} and 
a variety of variability patterns \citep{fenderbelloni04}. 
Its LFQPOs 
and their correlations with various other parameters have been studied 
extensively \citep[][etc.]{mac11,mik06, rodriguez02, muno99, morgan97}.
However, another intriguing aspect of this system is the spin
of the black hole. Various reports have found spin of \grs{} ranging 
from 0.56 \citep[][]{blum09} to $\sim$0.98 
\citep{mcclintock06,blum09}.
Probably, the most accurate spin estimate for \grs{} is available from the
\nustar{} observation \citep[][hereafter M13]{miller}, where the lower 
values of spin are ruled out at a high level of confidence based on the 
accurate modeling of relativistically blurred reflection. This 
observation belongs to a special state of \grs{} known as `plateau' 
state and it shows a QPO with frequency of 1.5 Hz. Properties of 
this LFQPO have been studied in detail by \citet[][hereafter Z15]{zhang} 
where the authors show that the RMS amplitude of the QPO increases
with energy with a characteristic flattening between 10$-$20 keV. 
This suggests the possibility of an interplay between independently 
varying spectral parameters. Thus investigation of the
spectral parameter(s), which lead to the observed energy dependence of the QPO power 
can provide significant clues on the origin of the QPO. 

In this context, here we delve into the energy dependence of the QPO
power by means of simulating light curves by varying different spectral 
parameter(s). In the next section, we discuss the \nustar{} observation 
and its analysis to reproduce the results of M13 and Z15. The results
of these spectral and timing analysis are used to generate the simulated
light curve as explained in the following section, with 
the discussion on our findings and preliminary conclusions in the 
subsequent sections.

 \begin{figure*}
 \centering
     \includegraphics[width=165mm,height=70mm]{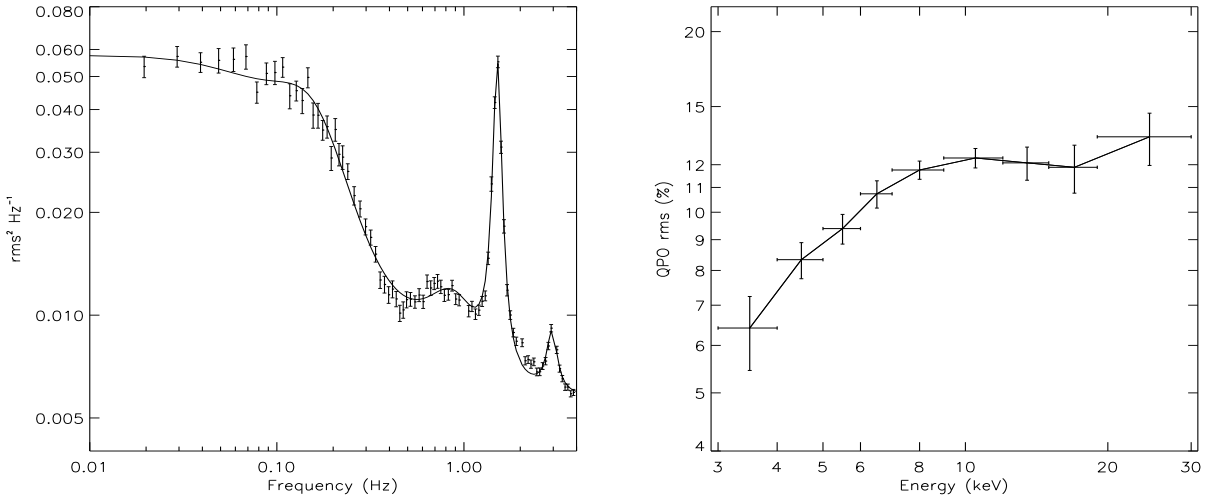} 
     \caption{Figure shows the observed PDS (left) in the energy range 
     of 3-80 keV and the energy dependence of QPO power (right).}
 \label{pds-spec}
 \end{figure*}

 \begin{center}
\begin{table*}
\centering
 \caption{The best fit spectral parameters calculated with the model {\tt tbabs$\times$(kerrconv$\otimes$reflionx{\_}hc + cutoffpl)}. 
 Errors indicate 90\% confidence intervals.}
    \begin{tabular}{p{3.5cm} c c }
    \hline
    \hline
Parameter     & Best fit value  \\
\hline
$N_H$ ($\times$10$^{22}$cm$^{-2}$) & 4.43$^{+0.18}_{-0.05}$\\
Index 1 			   & 9.99$_{-0.06}$\\
Index 2 			   & 0.01$^{+0.02}$\\
R$_{in}$ (R$_g$) 		   & 7.49$^{+0.07}_{-0.06}$\\
a$_\ast$			   & 0.973$^{+0.002}_{-0.002}$\\
$i$				   & 65.41$^{+1.02}_{-0.39}$\\
$\Gamma$ 			   & 1.709$^{+0.001}_{-0.001}$\\
$Xi$				   & 993.0$^{+14.8}_{-19.0}$\\
$A_{Fe}$			   & 0.97$^{+0.02}_{-0.13}$\\
Cut-off Energy			   & 35.47$^{+1.19}_{-0.18}$\\
$N_{ref}$ ($\times$10$^{-5}$)	   & 1.19$^{+0.20}_{-0.02}$\\
$N_{pl}$			   & 1.907$^{+0.008}_{-0.007}$\\

\hline

    \end{tabular} 
 \end{table*}

\end{center}

\section{Observation and Data Analysis}
The \nustar{} observation of \grs{} carried out on July 03, 2012 has 
been used for the present study. 
The same observation has been analyzed in detail by M13 and Z15 
for spectroscopic and timing studies respectively.
We extract the spectrum following M13 and adopt their best fit model, 
{\tt tbabs$\times$(kerrconv$\otimes$reflionx{\_}hc + cutoffpl)} 
where the reflection of the primary cut-off power law is modeled by 
{\tt reflionx{\_}hc} (a variant of {\tt reflionx} model), which assumes the 
incident spectrum as cut-off power law. 
With the same spectral model, we ensure that the best fit results 
are almost identical, which are then used in our simulations 
as explained in the next section.

We also carried out the timing analysis on the line of Z15.
The light curves are extracted in an energy range
of 3-80 keV with a bin size of 100 ms. 
PDS is generated for each interval of 1024 bins
and co-added for all the intervals. The PDS from both
FPMA and FPMB show a very strong type-C QPO at $\sim$1.5 Hz.
The PDS also shows strong spikes at every 1 Hz, however, it is known 
that these are instrumental artifacts in the early \nustar{} observations 
and hence we ignore these bins during fitting in {\tt XSPEC}. 
The light curves from FPMA and FPMB are added together in order to improve
the statistics and the PDS is generated from the combined light curve
(Fig.~\ref{pds-spec} left panel).
We fit the PDS using multiple lorentzians and a constant, which takes
care of white noise level in the PDS.
It is obvious that the PDS has many complex features such as broad band
noise, a small peak at 
frequency of $\sim$0.75 Hz and a strong QPO at $\sim$1.5 Hz. Here we concentrate
only on the $\sim$1.5 Hz QPO feature because it is known to have systematic
dependence on energy (Fig.~\ref{pds-spec} right panel).
The small peak at $\sim$0.75 Hz has been described as a sub-harmonic 
by Z15. However for the purpose of the 
present work, we consider it as a part of broad band noise mainly because 
of very low Q value of $\sim$3.6.
Z15 have reported that the RMS 
amplitude of the QPO increases with energy.
We verify this from the light curves in different energy bands (3$-$4 keV, 
4$-$5 keV, 5$-$6 keV, 6$-$7 keV, 7$-$9 keV, 9$-$12 keV, 12$-$15 keV and 15$-$30 keV).

We find an increasing trend with a shallow dip (or flattening) in 10$-$20 
keV as reported in Z15 (Fig.~\ref{pds-spec} right panel).
They have also reported that this observation shows increase in count rate
as well as QPO frequency during the later part (after $\sim$40 ks of the start 
of the observation). Therefore,
we restrict both the spectral and timing analysis to the first 40 ks of 
data, where the count rate and QPO frequency are quite stable. 
Effective exposure times for our analysis are 9.71 ks and 10.01 ks for 
FPMA and FPMB respectively.


\section{Light Curve Simulation and Results}
In many cases, the observed spectrum can be explained with different
models with same statistical significance. In such cases, the compatibility of a 
spectral model with the observed timing characteristic can be helpful to solve the 
degenracy.
The main objective of the present work is to investigate the compatibility 
of the observed QPO properties with the spectral model. One possibility to
achieve this is by comparing the amplitude of the oscillations across 
multiple energy bands. 
Another possibility is by comparing the phase lags
in a similar manner. 
Black hole binaries are known for showing phase lags between different 
energy bands. Phase lags for \grs{} has also been reported by several 
authors \cite[e.g][]{reig2000,qu10,lin2000}, wherein it is generally 
found that the hard photons lag the soft photons. We also calculated 
phase lags between different energy bands using \nustar{} data, however 
because of the relatively poor statistics (compared to e.g. RXTE-PCA), 
the phase lags can not be constrained (obtained values are consistent 
with zero with larger error bars) with the present data.
Hence we adopt the alternate method by comparing the observed energy 
dependence of the QPO RMS amplitude with the simulated energy dependence.

An important feature of the present best fit spectral model is that it 
consists of only one primary continuum component, cut-off power law, 
covering the full energy range. Hence, it does not allow interpretation 
of the QPO in terms of oscillation of one of the spectral component as 
suggested by \citet{rao2000} \citep[also corroborated by][]{vadawale01,
vadawale03}. Thus observed energy dependence of the QPO power must 
result from the variation of one or more spectral parameters. 

Here we assume that a particular spectral parameter varies with a 
frequency close to the observed QPO frequency.
We consider four possible parameters of the spectral model, viz. the 
spectral index, normalization of the cut-off power law, normalization 
of the reflection component and the ionization parameter; which can vary 
with the phases of the QPO. Though the high energy cut-off can vary, we 
keep it fixed at the best fit value because its variation does not affect 
the model in the energy range of interest. All parameters of {\tt kerrconv} 
are assumed not to vary with the phases of the QPO. 
We then simulate the light curve with a bin size of 100 ms. 
Initially we simulate the light curve for variation of only one parameter.
As a zeroth order approximation, we assume that the parameter is varying 
sinusoidally with QPO phases in a range centered at the best fit 
value.
The assumption of sinusoidal oscillation is clearly
an over simplification and the actual oscillation profile could be more
complex. However, the energy dependence of RMS amplitude is not likely to 
have strong dependence on the exact shape of oscillations.
We also include a small random variation over the sinusoidal variation to 
be a bit more realistic. 
Later we also simulate the light curve
for multiple parameters varying simultaneously. In either case the basic
algorithm for simulation of light curve is as follows-

It is assumed that a parameter $\alpha$ obtained from the best-fit 
spectral model varies at a frequency close to the observed QPO frequency 
($\bar{\nu}$). The amplitudes of the sinusoidal variation and random 
fluctuation are $p_\alpha$ and $q_{\alpha}$ respectively such that 
$q_\alpha$ = r~$p_\alpha$.
The simulation is carried out for r=0.1, 0.2 and 0.3 and it is found that 
the results remain unaffected. Therefore, we present results hereafter for 
r=0.1 (i.e. 10\% random variations superimposed over the sinusoidal 
oscillations).

\begin{figure*}
 \centering
     \includegraphics[width=165mm,height=80mm]{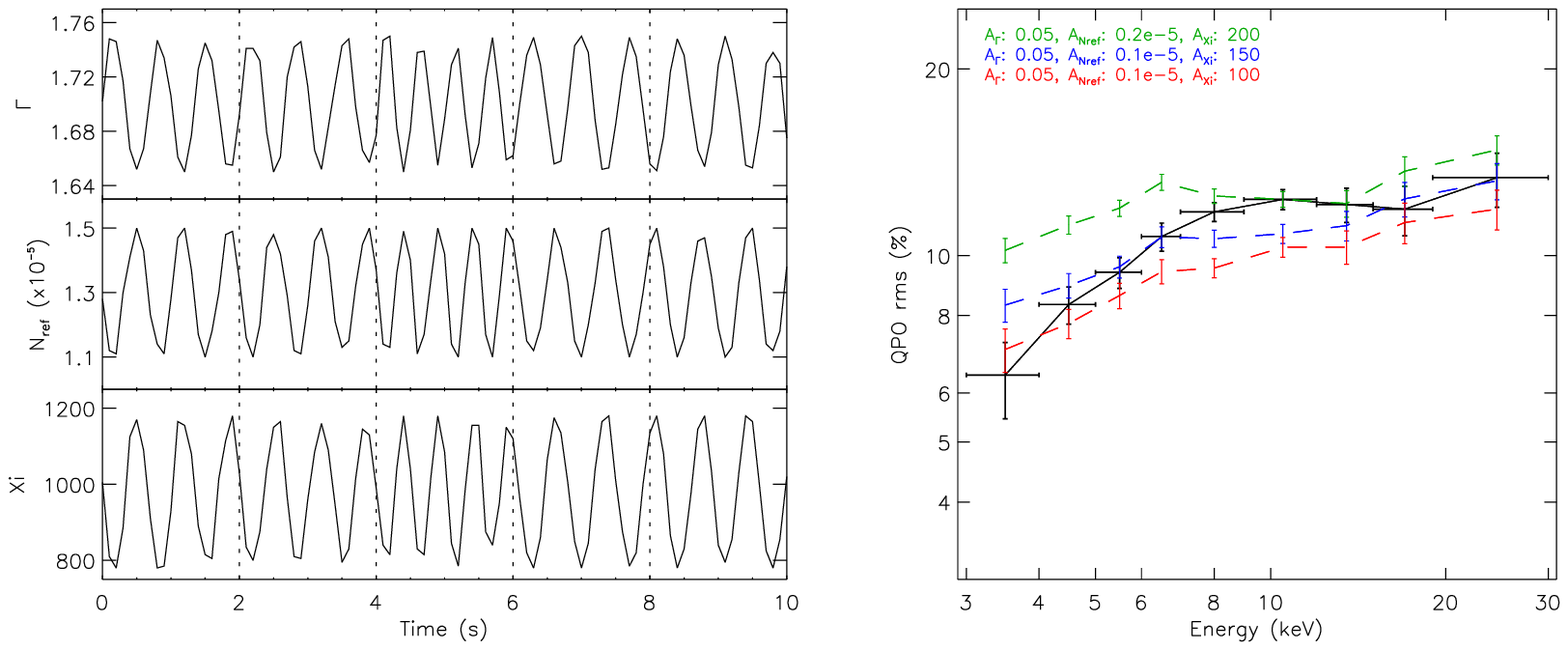} 
     \caption{The left panel shows the variation of three spectral parameters; spectral index ($\Gamma$), reflionx{\_}hc normalization ($N_{ref}$), 
     and ionization parameter (Xi). The reflection normalization
and ionization are varied in same phase which are assumed to be 
anti-correlated to the phase of spectral index variation. The dashed lines separate the intervals corresponding
to different frequencies of modulation. The right panel shows the simulated energy dependence of QPO power 
obtained by modulation of three parameters. Amplitudes of modulation for $\Gamma$, $N_{ref}$, and Xi 
are mentioned in the figure.}
 \label{simresult-3par}
 \end{figure*}

\begin{itemize}

\item
Let $m_{\alpha}$ be the best fit value of parameter $\alpha$.

\item
The spectral model is evaluated in {\tt XSPEC} with the parameter
$\alpha$ assuming values in the range
$m_{\alpha}\pm~p_{\alpha}$, keeping the rest of the parameters at their 
best fit values. Model counts, including the effects of detector response
matrix as well as effective area, are obtained in each of the energy bins 
and are tabulated.

\item
The light curve is simulated with a bin size of 100 ms for a duration of $\sim$ 10 ks to 
match the exposure time of the observation.

\item
For each interval ($\sim$ 2 s) a random frequency of oscillation ($\nu$) is drawn from 
a lorentzian distribution with the mean $\bar{\nu}$ (=1.5 Hz) and the
FWHM (=0.089 Hz) values same as those for the observed QPO. 
The frequency remains the same throughout an interval. 
We have also
verified that the overall results remain unaffected if the frequency 
is changed more often within one interval. It is assumed here that the 
quasi periodic nature arises due to frequency modulation \citep[see][]{fengyun10}.

\item
The phase of oscillation for each time bin $i$ is obtained by 
    $$ {\phi}_i = 2~\pi~\nu~{\Delta}t~+~\phi_{i-1}$$
where ${\Delta}t$ is the bin size and $\phi_{i-1}$ is the phase for previous 
time bin. Continuity of the phases across intervals is also maintained in the 
same manner. 

\item
The value assumed by parameter $\alpha$ for the $i$-th timebin is
calculated as  
$$ {\alpha}_i = p_{\alpha}~sin({\phi}_i) + q_{\alpha}~z_i$$
where $z_i$ is a standard normal variate
(see Fig~\ref{simresult-3par}, left panel).

\item
Mean counts in the time bin $i$ for a given energy bin $j$ ($\bar{C_{ji}}$)
corresponding to value $\alpha_i$ is obtained from the table generated in 
the first step. The energy bin $j$ is kept same as that used to extract the observed light curves.

\item
Random number is generated from Poisson distribution with mean 
$\bar{C_{ji}}$ and is recorded as counts in the $i$-th time bin of the 
light curve in energy bin $j$.

\item
Power density spectra are generated for these simulated light curves
using {\tt POWSPEC} and are fitted with {\tt XSPEC} as usual.

\end{itemize}

 \begin{figure*}
 \centering
     \includegraphics[width=150mm,height=170mm]{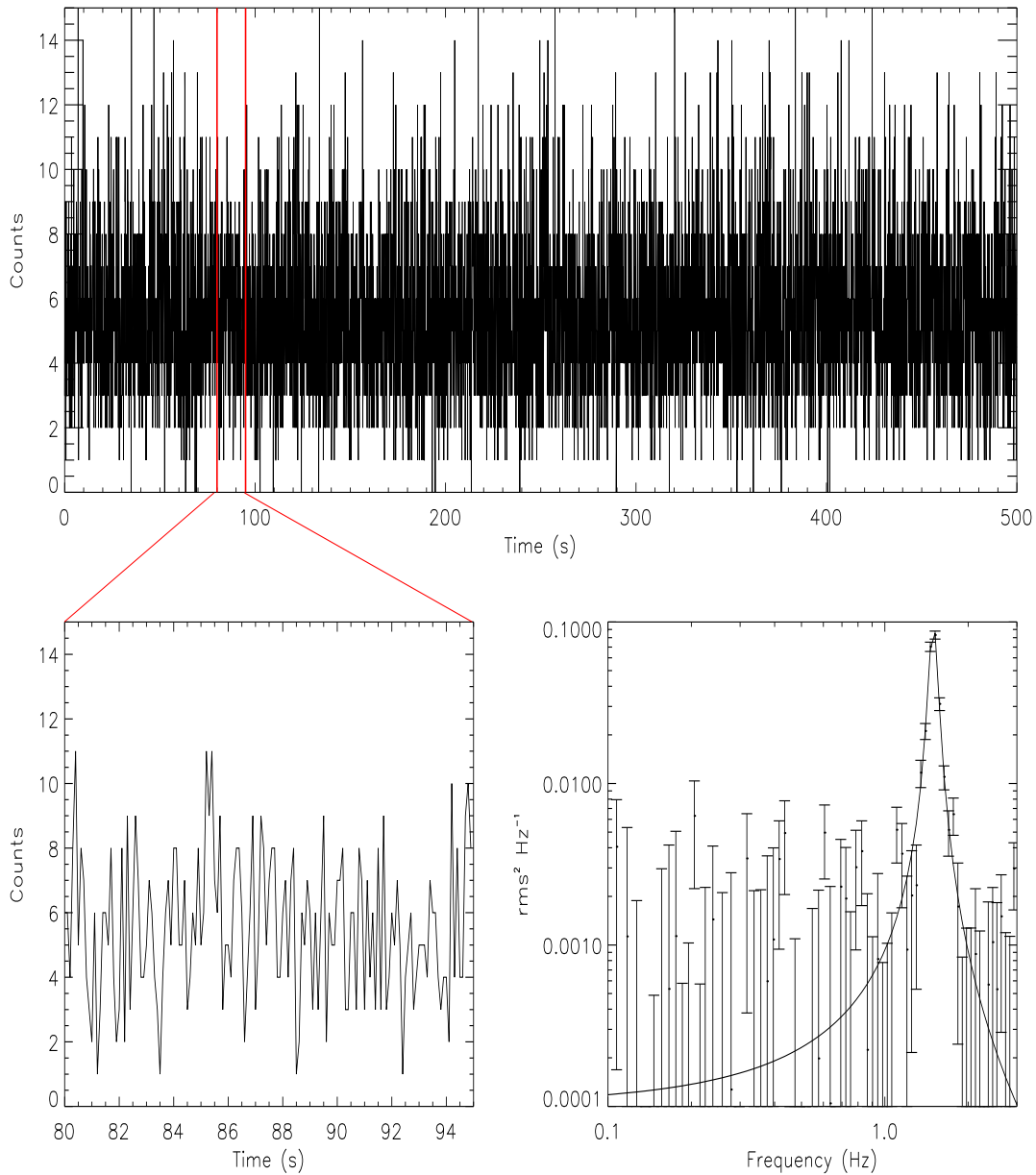} 
     \caption{The upper panel shows the first 500 seconds of a 9-12 keV light curve simulated by 
     modulating the spectral index. A zoomed-in view of a section of the light curve is 
     shown in the lower-left panel. PDS of the full light curve is shown in lower right panel.}
 \label{simlc}
 \end{figure*}

 \begin{figure*}
 \centering
     \includegraphics[width=160mm,height=130mm]{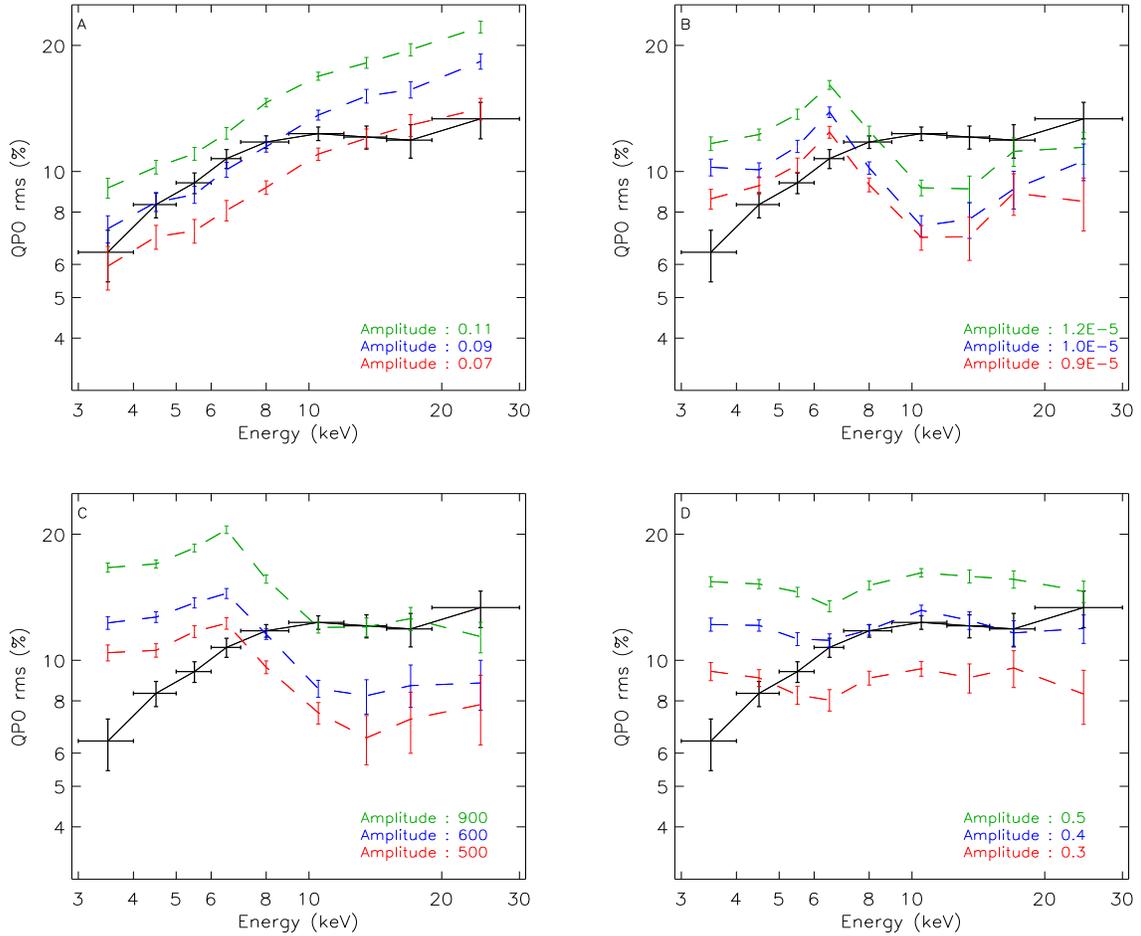} 
     \caption{Figure shows the simulated energy dependence of QPO obtained by modulation of spectral index
     (A), normalization of reflionx{\_}hc (B), ionization parameter (C) and normalization of cut-off
     power law (D). The dashed lines correspond to simulation results for different amplitudes of modulation.
     The observed energy dependence is shown with black solid line in all the panels.}
 \label{simresult-1par}
 \end{figure*}

  \begin{figure*}
 \centering
     \includegraphics[width=160mm,height=70mm]{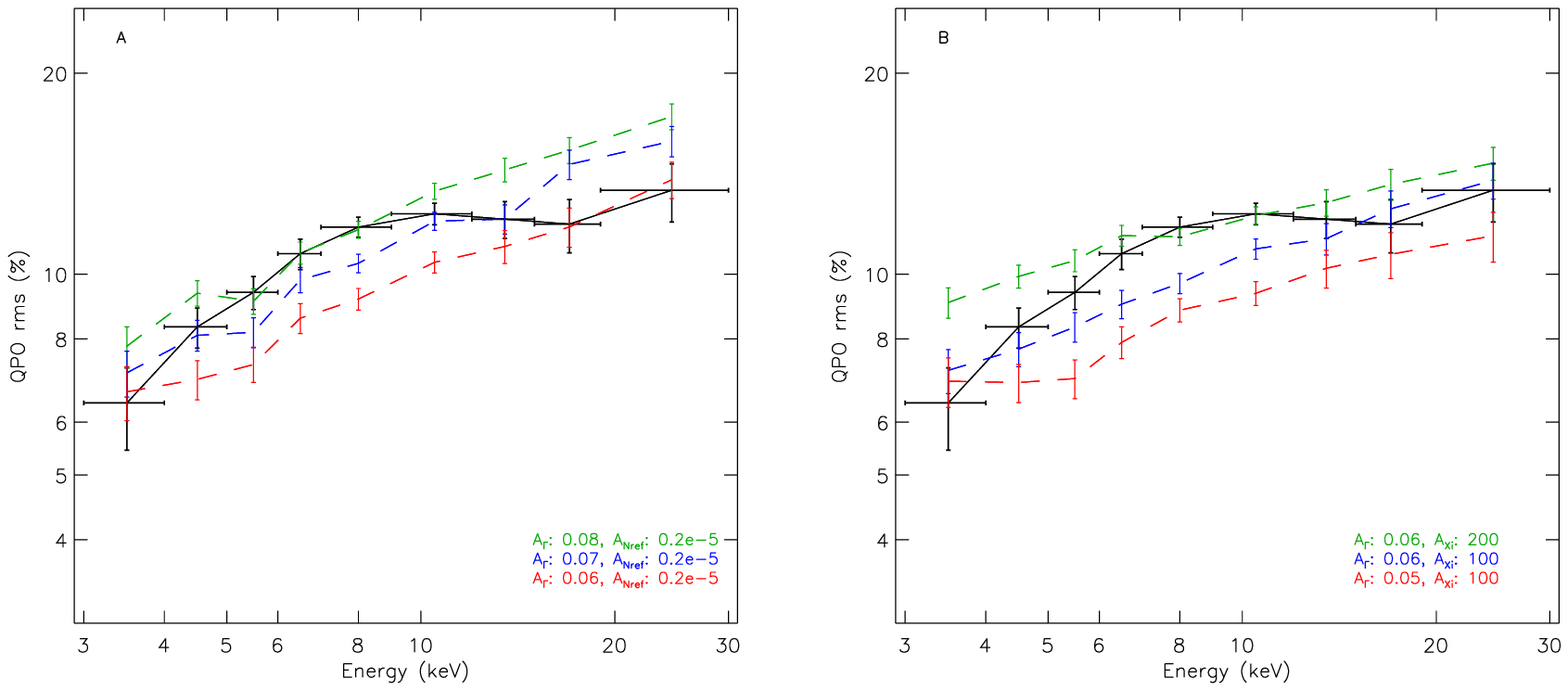} 
     \caption{Plot showing the energy dependence of QPO with the simulated light curves by varying 
     two parameter simultaneously. Panel A shows the results obtained from the variation of normalization 
     of reflection and spectral index. Panel B shows the result for the variation of ionization parameter 
     and spectral index. The dashed lines correspond to simulation results for different amplitudes of modulation.
     The observed energy dependence is shown with black solid line in both the panels.}
 \label{simresult-2par}
 \end{figure*}

Fig~(\ref{simlc}) shows a section of the simulated light curve and its 
PDS in an energy range of 9$-$12 keV. It should be 
noted that we have not included low frequency noise in our simulations 
because our primary objective here is to investigate the energy dependence 
of the QPO power. 
Fig~(\ref{simresult-1par}) summarizes the overall results of our simulations. 
The four panels show the simulated energy dependence of the QPO power for 
variation of the four spectral parameters. The observed energy dependence 
is shown with black solid line in all the panels. For each parameter, 
multiple lines are shown corresponding to the three different range of 
variation of parameters. 
It can be seen that only the variation 
of spectral index can reproduce the overall trend of the increasing QPO
power with energy. Variation of other three parameters cannot
reproduce the observed trend. Even in the case of the spectral index variation, 
the exact shape of the energy dependence curve, particularly the dip or 
flattening in the energy range of 10$-$20 keV is not reproduced.
This suggests that, while the variation of spectral index during 
QPO phases is essential it is not the only source of the variability. 
Hence we test the hypothesis by simulationg the light curves considering 
the variation of more than one parameter simultaneously. The simulation was 
repeated for the variation of two and three parameters and the results are 
shown in Fig~(\ref{simresult-2par}) and (\ref{simresult-3par}) respectively. 
Since the reflection component represents the Comptonization component 
reflected from the disk, it is expected that the former would be stronger 
for a harder spectrum. Or in other words, the reflection component should 
be anti-correlated with the spectral index. Therefore, the parameters of 
reflection component ($N_{ref}$ and $Xi$) were varied out of phase with 
spectral index in the simulation of light curves. 
Panel A of Fig~(\ref{simresult-2par}) shows the energy dependence obtained 
from the variation of reflection normalization ($N_{ref}$) and spectral
index and Panel B shows the results for ionization parameter ($Xi$) and 
spectral index.
A comparison of Panel A of Fig~(\ref{simresult-1par}) and 
Fig~(\ref{simresult-2par}) shows that the variation of spectral index 
and normalization of reflection reproduce the observed energy dependence 
more closely than the variation of spectral index alone. We also studied 
the energy dependence with the light curves simulated by varying 
1) normalization of cut-off power law and normalization of reflection 
component and 2) normalization of cut-off power law and ionization 
parameter. The variation of the two set of parameters does not reproduce 
the observed increasing trend of QPO power.

We have further attempted to reproduce the observed pattern by varying three
parameters simultaneously. 
The left panel of Fig~(\ref{simresult-3par}) shows the variation of spectral 
parameters with time. Representative results obtained from the variation of 
three parameters are shown in the right panel of Fig~(\ref{simresult-3par}). 
It can be seen that the results start resembling the observed pattern more 
closely. We avoid attempts to exactly reproduce the observed pattern 
as this will require detailed fine tuning of the specifics such as number 
of varying parameters, amplitudes of variation, phase lags etc. More 
importantly, a relook into the spectral model itself might also be 
necessary which is beyond the scope of this paper. 
The main objective here is to convey that the present best fit spectral
model is compatible with the observed trend of the increasing QPO power
with energy when the spectral index is assumed to be varying. However, 
it might be necessary to look beyond this model in order to reproduce 
the exact energy dependence of the QPO power.

\section{Discussion and Conclusions}

Increasing trend of the LFQPO power with energy is well known 
\citep[e.g.][]{tomsick01,rodriguez04,rodriguez08}.
The trend can also be seen in the results 
shown by \citet{zdzia05,sobol06}.
However, origin of this trend in terms of variation of the spectral
components has not been investigated so far. Here we attempt to 
explain the energy dependence of LFQPO observed during the first 
\nustar{} observation of \grs{}. We simulate the light curve for variation 
of various spectral parameters and compare the resultant energy 
dependence with the observed one. We find that only variation of the
spectral index reproduces the overall trend of the increasing 
QPO power with energy. Variation of any other parameter does not 
reproduce the observed trend. 
This has significant implications on the feasibility of various 
models proposed to explain the origin of the QPO. 
For example, the fact that the variation of reflection normalization does not 
reproduce the observed trend suggests that the QPO models based
on geometric modulation \citep{millerhoman05} may not be realistic.

Among the other existing models for the low frequency QPOs, the model 
based on Lense-Thirring precession is considered the most successful model. 
It successfully explains the overall shape of the PDS and the variation of 
QPO frequency with the changes in spectral hardness, which is further linked 
with the truncation radius of the inner disk \citep{ingdone10}. However, 
the energy dependence of QPO is not apparently clear.
The similar situation exists with a magnetohydrodynamic model for QPO 
based on the accretion ejection instability \citep{tagger, varniere, rodriguez02}. 
The shock oscillation model \citep{molteni,charabarti,garain} does produce the spectral 
variation and it could be compatible with the energy dependent QPO power.
However, the simulation results obtained from the present spectral model
cannot be directly compared with the predictions of 
shock oscillation model.
It might be necessary to perform similar 
simulation with appropriate model and then verify the results.

Another observation is the flattening of QPO spectrum at 10-20 keV which 
cannot be reproduced with variation of only spectral index. The flattening 
seems to be more common as it has been observed at many other occasions 
\citep[e.g.][]{tomsick01,rodriguez04,rodriguez08,zdzia05,sobol06}.
Similar feature is also found in few other sources \citep{li2013}.
We have verified that the same feature is found in the RXTE observations
of \grs{} during the `plateau' states. We fit the spectra during these  
observations with the present spectral model consisting of cut-off power law 
and relativistically blurred reflection. The best fit spectral parameters 
are similar to that found for the present \nustar{} spectrum. 
We performed similar simulations using RXTE spectra and could
reproduce the overall trend of increasing power of QPO. However, the flattening 
is not found in the simulation results.
The observed flattening 
might be an indicator of the presence of more than one continuum component 
as suggested by \citet{rodriguez04}.

Overall our simulation results can put significant constraints on the
modulation mechanism assumed in various present theoretical models
of QPO. 
Recently, \citet{axeldone15} have made an attempt to understand QPOs 
with frequency resolved spectroscopy.
We suggest that the QPO phase-resolved spectroscopy, similar
to that shown by \cite{ingklis15} can be another promising way to validate 
the spectral models, but with full energy resolution spectra. 
Recently launched Indian astronomy mission Astrosat \citep{astrosat}, 
with large effective area of the LAXPC instrument, will provide ample 
opportunities for such QPO phase-resolved spectroscopy owing to
its event mode data over broad energy range.

\acknowledgments
This research has made use of data obtained from High Energy 
Astrophysics Science Archive Research Center (HEASARC), provided by 
NASA's Goddard Space Flight Center. 
Research at Physical Research Laboratory is supported by Department of
Space, Govt. of India. We thank A. C. Fabian and M. Parker 
for providing {\tt reflionx{\_}hc} model. We acknowledge the anonymous
referee for his/her valuable comments.


\end{document}